%% file: main.tex
\documentclass[a4paper, amsfonts, amssymb, amsmath, reprint, showkeys, nofootinbib, twoside]{revtex4-1}
\usepackage[english]{babel}
\usepackage[utf8]{inputenc}
\usepackage{graphicx}
\usepackage[colorinlistoftodos, color=green!40, prependcaption]{todonotes}
\input{preamble}
\usepackage[pdftex, pdftitle={Article}, pdfauthor={Author}]{hyperref} % For hyperlinks in the PDF
\begin{document}
\title{
Multiplicity of Stable Attractors in Disordered Neural Models
}

\author{Raffaele Marino}
\email[]{raffaele.marino@butterflydecisions.com}% Your name
   % \affiliation{University of Florence, Department of Physics and Astronomy, Via G. Sansone 1 - 50019 Sesto Fiorentino (FI), Italy}
    \affiliation{Butterfly Decisions srl, Via dei Principati 74 - 84122 Salerno, Italy}
\author{Roberto Livi}
\email[]{roberto.livi@unifi.it}% Your name
\affiliation{University of Florence, Department of Physics and Astronomy, Via G. Sansone 1 - 50019 Sesto Fiorentino (FI), Italy}
\affiliation{Istituto dei Sistemi Complessi, CNR, Via Madonna del Piano 10 – 50019 Sesto Fiorentino (FI), Italy}
\author{Antonio Politi}
\email[]{a.politi@abdn.ac.uk}% Your name
\affiliation{Istituto dei Sistemi Complessi, CNR, Via Madonna del Piano 10 – 50019 Sesto Fiorentino (FI), Italy}
\affiliation{Institute of Pure and Applied Mathematics, Department of Physics, University of
Aberdeen, Aberdeen AB24 3UE,
United Kingdom}

\date{\today} % Leave empty to omit a date

\begin{abstract}
We show how large-deviation statistics allows one to obtain reliable estimates 
of the multiplicity of stable fixed-points in a model of neural ordinary differential equations previously employed in computational tasks.
The result is obtained by developing a suitable perturbative method in the amplitude of the disorder.
It turns out that for not-too-large coupling strengths there are no qualitative differences between the symmetric case, when
the dynamics is a purely gradient evolution, and the asymmetric case, when limit cycles and chaos can, in principle, arise.
The selection of this specific model is dictated by pedagogical reasons, but we are confident that the approach can be extended
to other many-degree-of-freedom dynamical models characterized by different classes of random coupling matrices. 

\end{abstract}

\keywords{Stable Attractors, Disordered Neural Networks, Order-Statistics, Large-Deviation}

\maketitle

{\sl Introduction} - 
Models of random neural networks defined in terms of first-order
ordinary differential equations have been recently recognized as
effective candidates for solving various computational problems. For instance, in \cite{Pap1} a continuous version of the
Sherrington-Kirkpatrick model has been investigated as a testbed of
the relations between topological and dynamical complexity. There, by tuning the variance $\sigma$ of the i.i.d. Gaussian
entries of the random coupling matrix among the continuous variables
(neurons), one finds a phase transition from a single equilibrium state
to a chaotic one. The main result of that study is that for any finite
number $N$ of neurons in the network the model
%, the region at the edge of chaos,
%i.e. close to the transition point $\sigma=1$, 
exhibits an exponentially
large number %with $N$ 
of equilibrium points.
Following quite different motivations, another class of these models,
known as  Coherent Ising Machines (CIM's), has been studied as
nonconventional architectures for finding approximate solutions of
large-scale combinatorial optimization problems (e.g. \cite{Ghim, Berl1,
Berl2}). In CIM's the continuous neural variables are subject to a local
double-well potential and they are coupled via a symmetric matrix,
whose random entries are i.i.d. Gaussian variables. Due to its gradient-dynamic structure, CIM's are typically
employed for identifying global energy minima by making use of an
annealing process, starting from a
single trivial ground state and evolving towards stable fixed points.
Another model, inspired by Recurrent Neural Networks \cite{Senjowski1,
Senjowski2} and much similar to CIM's, is the set of neural Ordinary 
Differential Equations (nODEs), recently analyzed in \cite{Pap2}.
It again deals with
continuous neural variables in a double-well local potential. 
However, the coupling matrix is non-symmetric, thus allowing
for a chaotic evolution in the limit of large coupling values. 
Such a model has been employed in a region of parameters space
containing spontaneous or planted stable fixed-points.
This dynamical system has been found able to perform standard learning
tasks, making use of implicit feed-forward modules \cite{Pap2}.
The procedure exploits the presence of
large sets of stable attractors (typically, fixed-points), 
employed as targets of a learnable dynamics, where the Euler
numerical integration outlines the recurrent architecture of
deep-learning algorithms. Moreover, it has been shown that
effective training strategies can be applied to enforce the access
to planted attractors, representative of classification patterns
\cite{Pap4, Chicchi2025, MarinoWC2025}.
All of these models were investigated having in mind specific
computational tasks, but no systematic effort has been devoted
to understanding the underlying dynamical structure, common to this wide class
of models of nODEs.

In this Letter we provide a preliminary account of such aspects,
by focusing our attention on the dynamical and statistical complexity
of the model studied in \cite{Pap2}.
More precisely, we describe how the exponentially large
number, $2^N$ ($N$ being the system size), of stable fixed points 
present for zero coupling, $g=0$,
decreases, when $g$ is switched on. An
exponentially large number of stable fixed-points survive, at
least up to some finite value of $g \sim {\mathcal O}(1)$.
In this range of $g$ values we construct
a large-deviation functional, by exploiting the extreme-value
statistics, associated with the fixed-points depletion mechanism. 
All of this is achieved thanks to a perturbative criterion.
To our knowledge, this is a fully novel strategy, which discloses yet unexplored
perspectives in relating many-degree-of-freedom dynamical systems with the
statistics of disordered models, such as spin-glasses.
%We shortly discuss how this criterion applies also when different classes of random coupling matrices are considered.
Final remarks about the dynamics in the large coupling limit envisage future
studies about the dynamical phases of the model, characterized by the presence
of stable limit cycles and chaotic attractors.
\vskip .2 cm
{\sl The model} - We study a set of $N$ \textit{particles/neurons} $x_i$, which satisfy the
following dynamical equations
\begin{equation}
\label{dyneq}
\frac{d x_i }{dt} = - x_i(x_i^2 -1) + \frac{g}{\sqrt{N}} \, \sum_{j = 1}^{N} \, A_{ij} x_j \, .
\end{equation}
Each particle is subject to a local double-well potential, whose minima are located at $\pm 1$ and separated by
a maximum in $0$, namely $V(x_i) = \frac{1}{4} (x_i^2 - 1)^2$.
Additionally, the particles mutually interact via an $N \times N$ random matrix $\mathbf{A}$, while
$g$ denotes the coupling strength.
The entries $A_{ij}$ follow a standard Gaussian distribution $\mathcal{N}(0,1)$, so that
$\mathbf{G} = \frac{1}{\sqrt{N}}\,\mathbf{A}$ is an element of the \emph{real Ginibre ensemble}.
%Since $\mathbf{G}$ is, by definition, not symmetric,
%\emph{Girko's circular law} tells us that, as $N\to\infty$, the eigenvalues
%of $\mathbf{G}$ concentrate in the complex plane inside the unit disk.
%Finally, g
Given the symmetry of the nODEs, dynamics (\ref{dyneq}) is invariant under the change of sign, i.e.
$ \{ x_i \} \rightarrow  \{ - x_i \} $.

For $g=0$ the dynamics reduces to a Cartesian product of overdamped processes.
Given a generic initial condition, each $x_i$ moves to its closest minimum and
the overall dynamical system collapses onto one of the $2^N$ stable fixed
points (SFPs), $\{\vec{x}^{*(k)}(0)\}$ with $1 \le k \le 2^N$,  corresponding to all sequences of $\pm 1$.
As soon as the coupling $g$ is switched on, mutual interactions arise
and it is natural to expect deviations of the coordinates
of the SFPs from the initial $\pm 1$ values.

%yield a dynamics where any initial condition
%evolves to a slightly perturbed fixed point among those at $g=0$.
%In fact, the effect of the coupling term can be thought of as a random
%perturbation of the double well potential, whose shape and minima
%are subject to very small modifications.
%Upon further increasing $g$, qualitative changes emerge, which include
%tangent bifurcations, emergence of limit cycles and even chaotic attractors.

\begin{figure}[t]
  \centering
  \includegraphics[width=1.\linewidth]{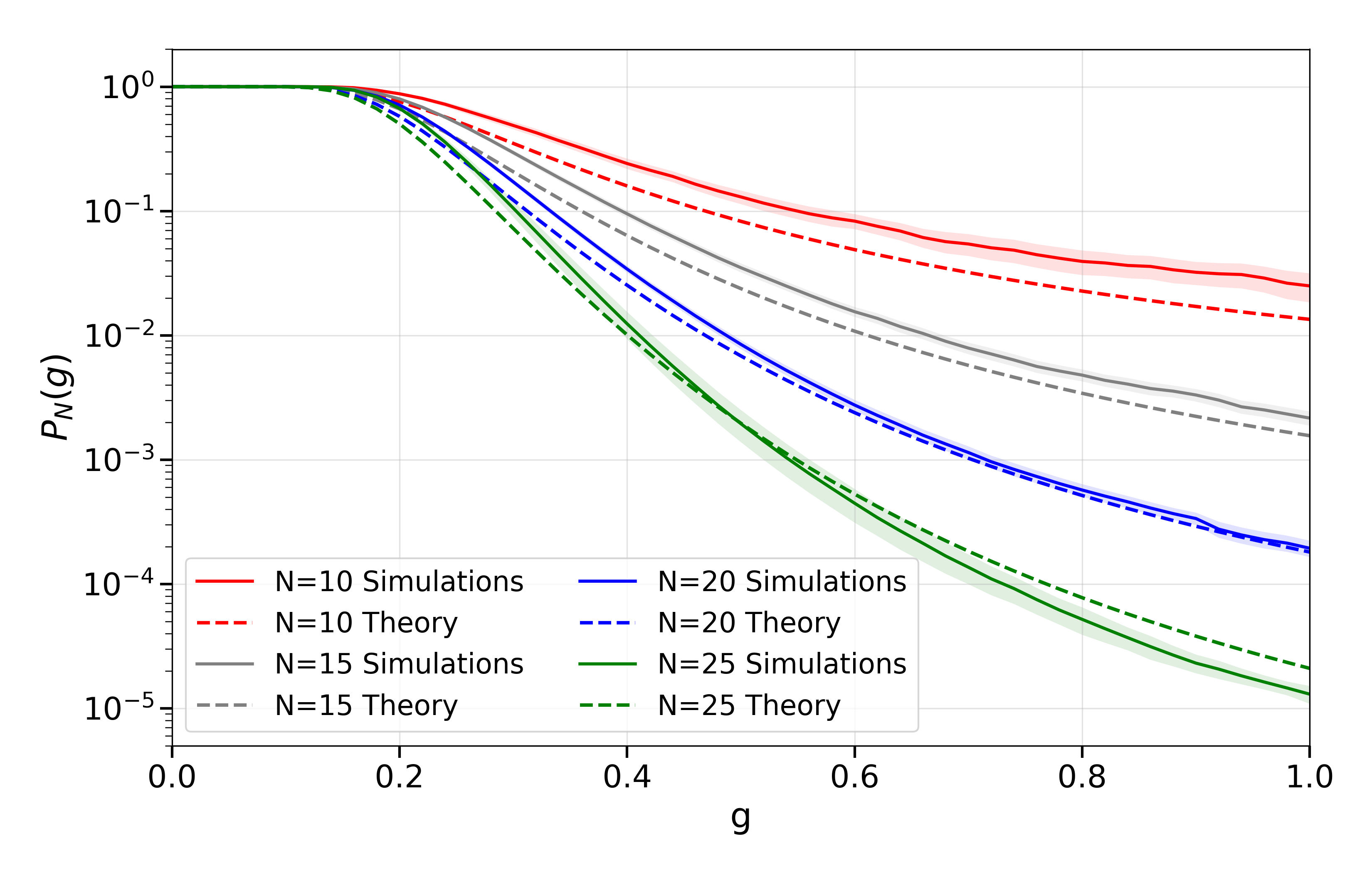}
  \caption{Fraction of $P_N(g)$ of SFPs as a function of $g$ for several sizes.
  Solid curves are simulation estimates averaged over different matrices ($10$ for $N\leq 20$, $2$ for $N= 25$); shaded bands indicate the uncertainty on the mean (error bar on the average).
  Dashed curves are the perturbative predictions obtained by integrating the minima density.}
  \label{fig:smallN_fixedpoints}
\end{figure}

\vskip .2 cm
{\sl Results} -
To build intuition on the qualitative organization of the dynamics, we first
analyze the system for small sizes $N$. In this setting, the attractor
landscape can be examined systematically as a function of the coupling
parameter $g$.
Since we expect the number of SFPs to depend on the realization of the matrix,
the data reported in Fig.~\ref{fig:smallN_fixedpoints} are obtained by averaging
over different matrices.
There we report the average fraction $P_N(g)$ of
SFPs as a function of $g$, for several values of $N$.
For small $g$, the curves remain close to $1$ suggesting that
the number of SFPs is constant.
Upon increasing $g$, a significant drop (notice the logarithmic vertical
scale) is observed, and the decrease becomes progressively steeper as $N$
grows.
Beyond this drop, there are much fewer SFPs, and for $g > 1$ limit cycles as well as
chaotic attractors may appear (data not reported).
So long as SFPs persist, we have observed that they retain the same
\emph{sign code}: namely, the vector
\begin{equation}
\label{signeq}
    \vec{s}^{(k)}(g) \;=\; \mathrm{sign}\!\big(\vec{x}^{*(k)}(g)\big)\in\{\pm 1\}^N
\end{equation}
coincides with the sign pattern present at $g=0$.
In this sense, the surviving fixed points can be interpreted as
\emph{descendants} of the SFPs at $g = 0$, even though the values of
their components change with $g$.
\begin{figure}[t]
  \centering
  \includegraphics[width=1.\linewidth]{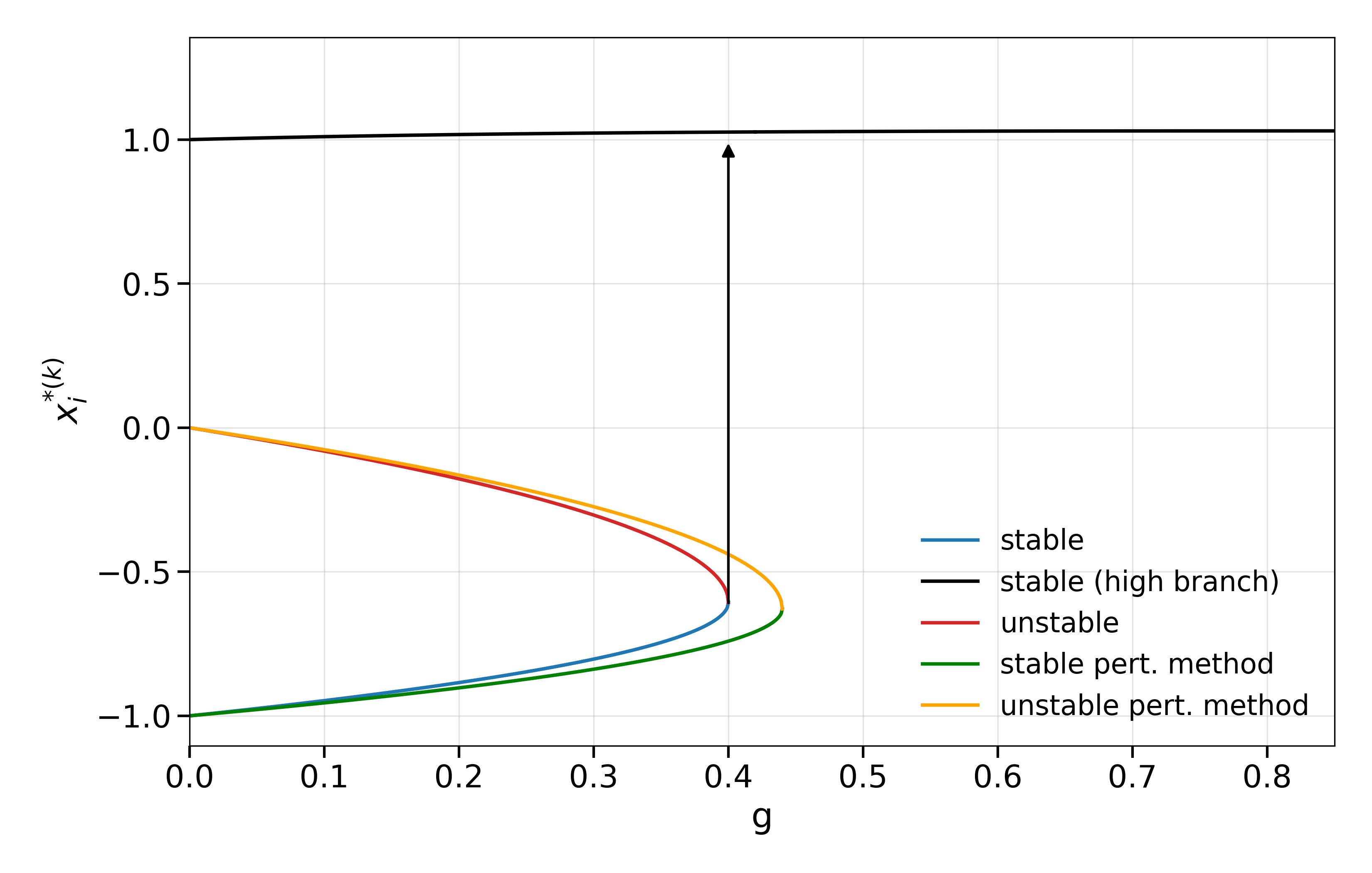}
  \caption{For small $g$, the stable fixed point (blue, lower branch) is a descendant of the $g=0$ sign pattern with $x_i^{*(k)}\approx -1$, and the component drifts continuously as $g$ increases.
The unstable equilibrium branch (red) approaches the stable one and annihilates with it near $g\simeq 0.4$. %, consistent with a tangent (saddle--node) event where the stable and unstable equilibria collide and disappear.
%For $g>0.4$, trajectories that previously converged to the vanished fixed point are instead captured by a different stable fixed point (black upper branch, with $x_i^{*(k)}\approx +1$). 
In green (orange) is presented the stable (unstable) path computed using our perturbative method. }
  \label{fig:tangent_component_N10}
\end{figure}

The depletion of SFPs is due to a sequence of \emph{tangent} (also called
\emph{saddle-node} or \emph{fold}) bifurcations.
In fact, for $g=0$ there exists a much larger number of unstable fixed points
(UFPs) $\{\vec{y}^{*(k)}(0)\}$, usually {\sl saddles} of different order, corresponding to
the $(3^N - 2^N)$ length-$N$ sequences of 0s and $\pm 1$s,
containing at least one 0. For each SFP there exist $N$ neighbouring UFP's,
obtained by setting one of its components equal to 0.
Analogously to the SFPs, saddles can be continued analytically starting from
$g=0$.
Upon increasing $g$, it may happen that one SFP $\vec{x}^{*(k)}(g)$
approaches one of the neighbouring saddles $\vec{y}^{*(l)}(g)$ and eventually
the two points mutually annihilate for some SFP-dependent critical value $g_c$.
This scenario is clearly illustrated in Fig.~\ref{fig:tangent_component_N10},
where one component of an SFP is reported as a function of $g$. The blue line
corresponds to the SFP; beyond $g_c\approx 0.4$ an initial condition in the
vicinity of the no-longer existing SFP converges towards another SFP, which
inherits the basin of attraction of the disappeared SFP.
The disappearance of the SFP could be equally monitored by following any
component of the SFP, since the bifurcation is a general phenomenon, which
implies that all components of the two colliding fixed points simultaneously
collapse with one another. 
%An alternative method for identifying these
%saddle-node bifurcations is shortly illustrated in Appendix \ref{app:Jacobian}.
We conjecture that the bifurcation is driven by a single component, the
most sensitive to the coupling.

The bifurcation mechanism is well captured by a perturbative approach.
Given an SFP at $g=0$ with components
$x^*_j(0)\in\{\pm 1\}^N$, we introduce the induced field
(see Eq.(\ref{signeq}) ) $S_i \;=\; \frac{1}{\sqrt{N}}\sum_{j=1}^N
A_{ij}\,s^{(k)}_j$,
and thereby approximate the dynamics of the $i$-th component as
$\dot x_i = x_i(1-x_i^2) + g\,S_i$.
In this approximation, the evolution still factorizes into the
independent relaxation of the various components within quartic potentials.
For a given $g$, an SFP exists so long as its components correspond to a local
minimum.
Since a positive (negative) $S_i$ tends to destabilize a negative (positive) $s^{(k)}_i$, 
it is convenient to introduce
$u_i = S_is^{(k)}_i$ (for the sake of simplicity we drop the index $k$).
Let us now denote with $i$ the component of the SFP characterized by
the most negative $u_i$ and be $u$ its value.
Given $u$, it is readily seen that the critical value where such a minimum
disappears is
\begin{equation}
g_c\;=\;-\frac{2}{3^{3/2}\,u},
\label{eq:ug}
\end{equation}
The accuracy of this approach can be appreciated in Fig.~\ref{fig:tangent_component_N10},
where the $j$-th component (therein $i=1$) of an SFP determined via the perturbative
approach (see the green line) can be compared with the actual exact solution. The two
curves are very close to one another up to the critical point, in spite of $g_c$ being not too small.
The advantage of this approximate approach is that one can infer the range of existence
of the various SFPs directly from the knowledge of the matrix $A$, without the need of
performing simulations for different coupling strengths. In fact, if we denote with
$\rho_N(u)$ the normalized empirical density of $u$ values
(averaged over disorder realizations), we can express the fraction $P_N(g)$
of SFPs for a given value of $g$, as $P_N(g)=\int_{u_c(g)}^{+\infty}\rho_N(u)\,du $
where $u_c(g)= -2/(3^{3/2}g)$ is obtained by inverting Eq.~(\ref{eq:ug}).
\begin{figure}[t]
  \centering
  \includegraphics[width=0.9\linewidth]{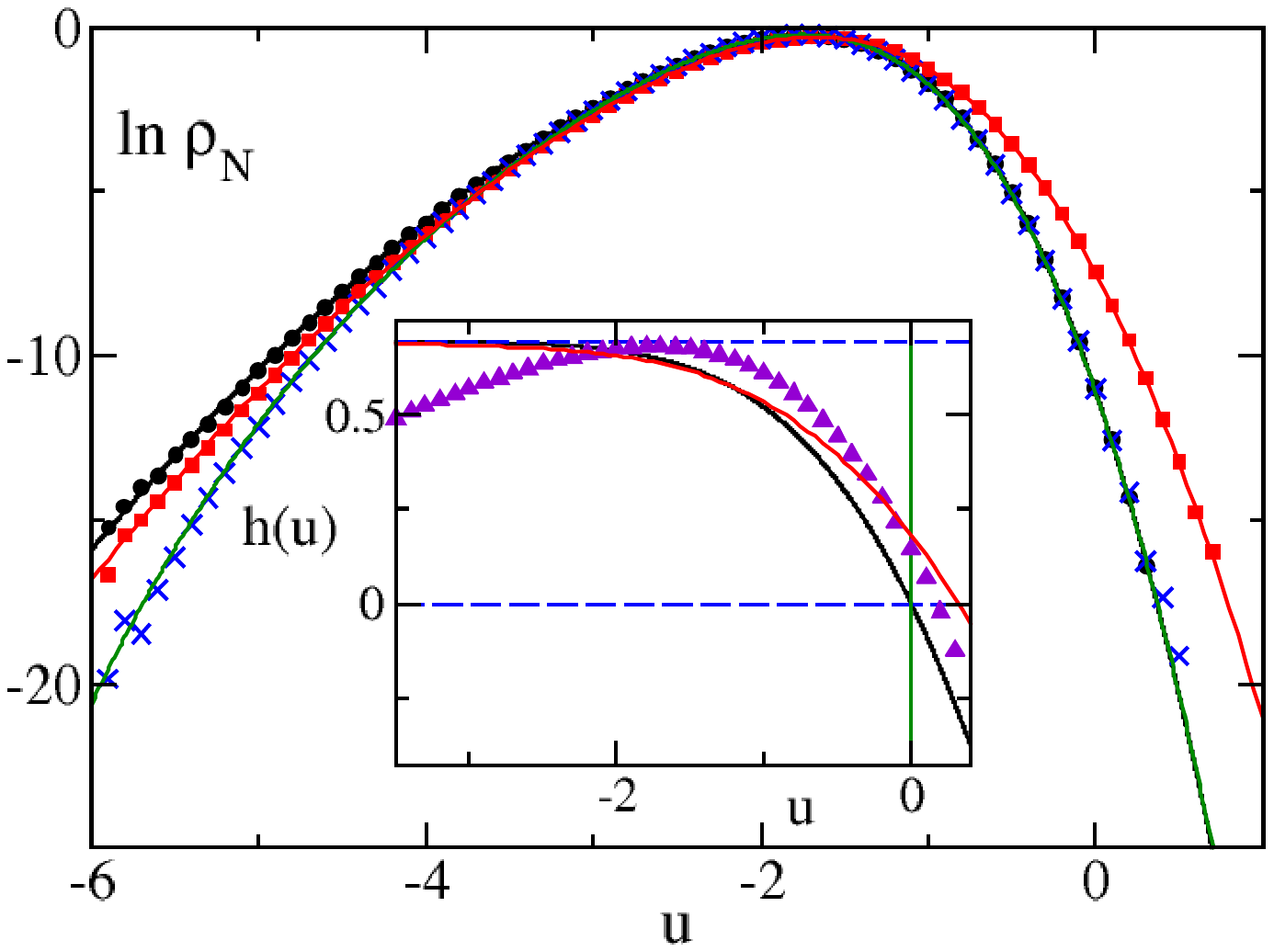}
  \caption{Main panel: the symbols denote numerical estimates of $\rho_N(u)$ for $N=20$. Data have been obtained by averaging over
  1000 realizations of the matrix $\mathbf{A}$. Full circles, crosses, and squares correspond to the Ginibre ensemble, a
  uniform distribution of entries, and to the symmetric case, respectively. The solid curves identify the corresponding
  theoretical predictions.
  Inset: the triangles correspond to $h(u)$ estimated as $\ln \rho_N/N+ \ln 2$ for the Ginibre ensemble. 
  %  , while solid black circles have been obtained from the logarithmic derivative of $\rho_N(u)$, comparing $N=23$ with $N=20$ and by adjusting the vertical shift to eliminate
% residual finite-size effects. Diamonds have been obtained for symmetric matrices.
The black and red solid curves correspond to the theoretical predictions for asymmetric and symmetric matrices, respectively.}
  \label{fig:scaling}
\end{figure}
The resulting fractions $P_N$ correspond to the dashed line in Fig.~\ref{fig:smallN_fixedpoints}:
they agree pretty well with the direct numerical data over several decades, confirming the
validity of the perturbative scheme.

Since $\rho_N(u)$ contains the relevant information
to characterize the stability of the various SFPs, we now focus on its structure.
The numerical estimates for $N=20$ are reported in Fig.~\ref{fig:scaling} (see full circles).
Notice that positive $u$ values are irrelevant in the context of the stability analysis,
since they correspond to SFPs which become more, rather than less, stable,
under the action of the mutual coupling. Moreover, given the inverse proportionality between
$u$ and $g$, the very negative $u$ values identify
the most fragile SFPs: those which disappear for very small coupling strengths.

Given the matrix $\bf A$, $u$ is the minimum value within
a set of $N$ variables $S_i$, each one obtained by summing $N$ independent elements,
multiplied by {\it de facto} random independent signs
(we sum over all SFPs and hence all sign patterns).
Therefore, we can invoke order-statistics identities for generic distributions
(see  \ref{app:minima_mean_var}).
In particular, by denoting with $\varphi$, the probability density function (PDF) of the
$S_i$ elements and with $\Phi$ the corresponding cumulative distribution function (CDF),
it is known \cite{ross2020first} that
\begin{equation}
\rho_N(u) = N\,\varphi(u)\,\big(1-\Phi(u)\big)^{N-1}.
\label{eq:rho_exact}
\end{equation}
In the present case, $\phi(u)$ is the unit-variance Gaussian.
There is only a doubt on the validity of general theorems. Since the minimum is
taken after multiplying the variables $S_i$ by the pattern of signs employed in the
definition of $S_i$ itself, we cannot exclude subtle correlations sneak in.
Hence, we have directly tested  the validity of Eq.~(\ref{eq:rho_exact}).
The comparison can be appreciated in Fig.~\ref{fig:scaling}: the
theoretical prediction is indistinguishable from the
direct numerical results. The same is true for other values of $N$ (data not shown).

It is natural to invoke the large-deviation theory and conjecture that $\ln \rho_N$ is
asymptotically proportional to $N$; in mathematical terms,
$\psi(u) = \lim_{N\to\infty} (\ln \rho_N(u))/N$.
This is indeed correct, and from Eq.~(\ref{eq:rho_exact}), we see that $\psi(u) = \ln (1- \Phi(u)) $.
In fact, it is more instructive to refer to the actual (average) number
$E_N(u)=2^NP_N(u)$ of expected SFPs, rather than to their fraction.
Their growth rate is $h(u)= \ln 2 + \psi(u) = \ln (2(1-\Phi(u)))$. The resulting distribution
is shown in the inset of Fig.~\ref{fig:scaling} (see the black solid curve).
The triangles are obtained directly from  $(\ln \rho_N)/N$ (see the triangles): the initial 
rising part is a clearcut finite-size effect.

\begin{figure}[t]
  \centering
  \includegraphics[width=1.\linewidth]{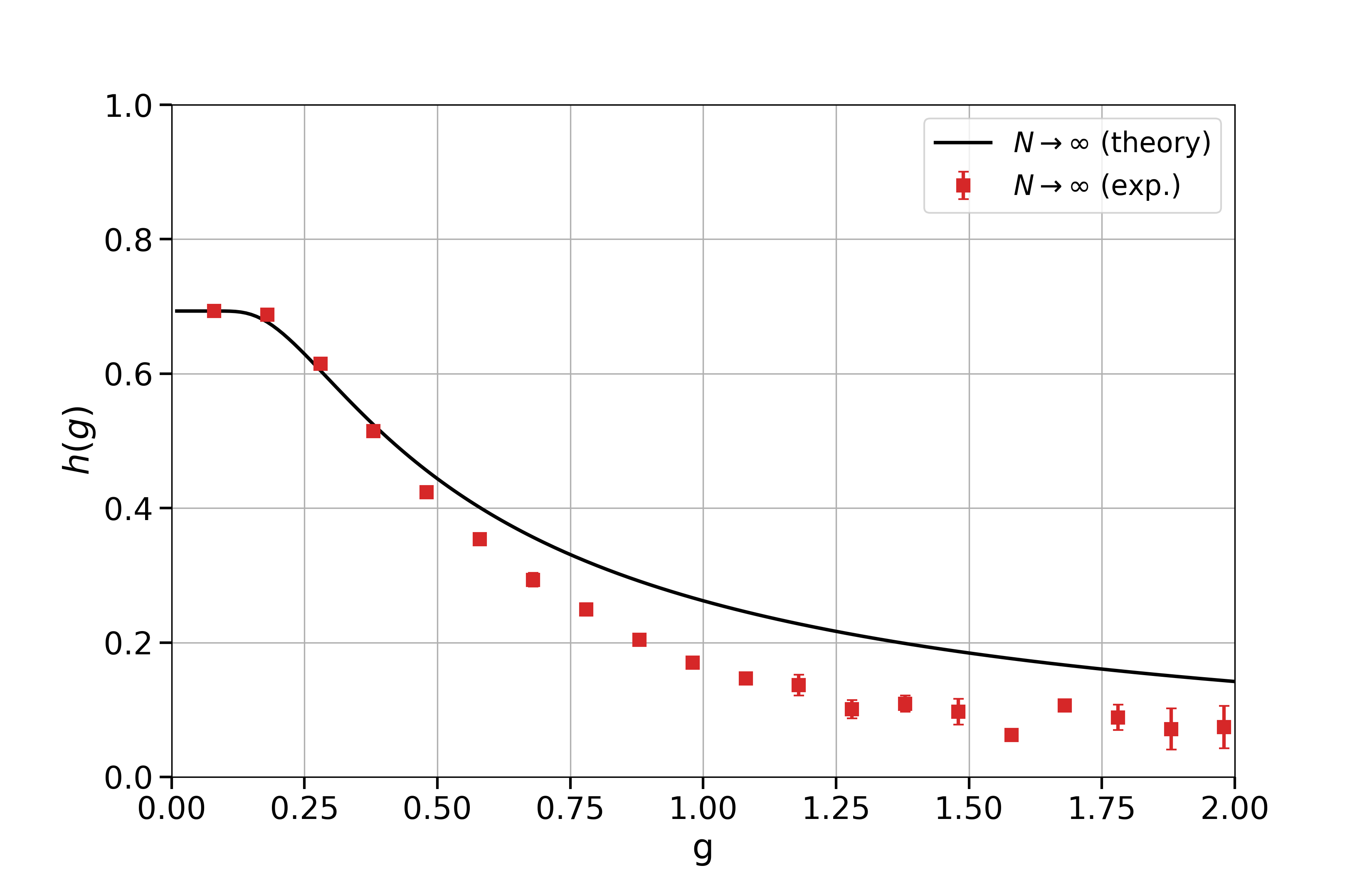}
  \caption{  Comparison between the perturbative theoretical prediction $h(g) = \ln 2 + \ln[1-\Phi(u_c(g))]$ (black solid line) and the rate extracted from direct numerical simulations in the large-$N$ limit (red squares with error bars).}
  \label{fig:hg}
\end{figure}
Finally, we have plotted the multiplicity index $h$ as a function of the coupling
strength $g$ (via Eq.~(\ref{eq:ug})) to allow for a comparison with direct
numerical simulations. The solid curve in Fig.~\ref{fig:hg} corresponds to the
perturbative estimate of $h(g)$, which is
strictly larger than $0$ for arbitrarily large $g$ values.
This is a consequence of the implicit assumption that the only way SFPs disappear
is via tangent bifurcations, which let anyhow $N$ SFPs survive. 
This is not true, as we have spotted various homoclinic and heteroclinic bifurcations,
which lead to either the disappearance or the destabilization of some SFPs (eventually to
the onset of limit cycles and chaos).
Whether or not these additional mechanisms can lead to a vanishing or even possibly negative $h(g)$
for a finite coupling strength is hard to say.
Direct numerical estimates of $h(g)$, obtained via a fit in the range $N\in[10,25]$, are in fact consistently
smaller than the theoretical prediction (see the red squares) and the difference becomes substantial for
$g$ larger than 0.5. However, it is appropriate to notice that numerical data are unavoidably
affected by finite-size corrections, quite difficult to quantify.

{\sl Different ensembles} -
If the matrix entries do not follow a Gaussian distribution, the central-limit
theorem anyhow implies that, in the large $N$ limit, $u$ should still be distributed in a
Gaussian way, though with likely deviations in the tails.
The relevance of such deviations, can be appreciated in Fig.~\ref{fig:scaling} (see the blue
crosses), where we report data obtained by assuming a uniform distribution 
in the interval $[-\sqrt{3},\sqrt{3}]$ (still unit variance) and $N=20$.
Appreciable differences can be seen only for $u$ smaller than $-4$ (i.e. $g < 0.1$).

Symmetric matrices represent an important class of models,
characterized by a strictly gradient dynamics, which implies
that the evolution can only converge onto an SFP.
Since the symmetry $A_{ij}=A_{ji}$ introduces strong correlations among the matrix entries,
we do not expect our theory to reproduce exactly the multiplicity of SFPs.
This is confirmed by the refined perturbative arguments developed in \ref{symmatr}.
However, as shown in Fig.~\ref{fig:scaling}), the numerical data obtained for
$N=20$ (see the red squares) do not differ significantly from those for the Ginibre ensemble.

However, an important qualitative discrepancy can be appreciated in the inset, where
one sees that the exponential growth rate $h(u)$ remains strictly finite for
$u\to 0$ (i.e. $g \to \infty$). Since in the large-$g$ limit, there is no guarantee that the perturbative approach 
provides sufficiently accurate results, we have performed direct simulations of
the full model for increasing $N$ and $g=5$. The data reported at the end of \ref{symmatr}
reveal a clear exponential growth, while no evidence can be found in the asymmetric case,
where periodic cycles and chaotic attractors {\it eat} a large fraction of the phase space.

{\sl Conclusions and perspectives} - In this Letter we have focused our attention on the survival mechanism of stable fixed points in model (\ref{dyneq}), showing that a suitable perturbative approach allows for reliable estimates,
making use of large-deviation theory.
We have also shown that such a perturbative approach is effective for different classes of random
coupling matrices, including symmetric ones. Preliminary results (to appear in a forthcoming
publication) indicate that the perturbative method is effective also when the random coupling matrix
contains planted attractors or has been passed through a training procedure for solving specific
computational tasks. In these cases, the matrix entries cannot be assumed anymore to be i.i.d.
variables and the standard order-statistics (see \ref{app:minima_mean_var}) does not apply.
In fact, the probability distributions $\phi(x)$ obtained from ensembles of the above mentioned
matrices typically acquire long-tails, testifying at their intrinsic non-Gaussian nature
and to the presence of correlations among the matrix elements.
However, we can argue that the perturbative method still applies to {\sl sum of variables}
and this is why it works pretty well, although providing less accurate estimates of the survival
probability of SFP's.
In the limit of large coupling strengths, preliminary studies show the appearance of limit cycles and
even chaotic attractors (see also \ref{app:largeg}).
Their relation with the survival of SFPs and with the presence of unstable
fixed points is quite an intricate problem. In order to tackle it successfully, statistical and
dynamical concepts and tools have to be suitably combined. For instance, in a more elaborated model
than (\ref{dyneq}) it has been found that in the ferromagnetic and paramagnetic chaotic phases there
is no direct correspondence between the presence of an exponentially large number of
unstable fixed points and the attractor manifold of these phases
\cite{FPRU}.

\bibliographystyle{unsrtnat}
\bibliography{references}

\newpage
\section{END MATTER}
\subsection{Mean and variance of the minima distribution}
\label{app:minima_mean_var}

Here, we summarize analytical formulae about the distribution of the minima
\begin{equation}
    u \equiv \min_{1\le i\le N} S_i
\end{equation}
that enters the perturbative scheme.
The key point is that, for a stable fixed point at $g=0$, the $S_i$ are i.i.d. Gaussian variables.
As a consequence, $u$ is the minimum of $N$ i.i.d. standard normal variables, whose distribution,
mean, and variance are controlled by classical order-statistics and extreme-value theory.

\subsubsection{From Ginibre to i.i.d. Gaussians}
Let $\mathbf{G}\in\mathbb R^{N\times N}$ be a real Ginibre matrix with
$A_{ij}\stackrel{\mathrm{i.i.d.}}{\sim}\mathcal N(0,1)$, i.e.,  $\mathbf{G}=\mathbf{A}/\sqrt N$.
For a stable fixed point at $g=0$ we set $\vec{s}\in\{\pm1\}^N$ and define
\begin{equation}
S_i \;\equiv\; \sum_{j=1}^N G_{ij}s_j
\;=\;\frac{1}{\sqrt N}\sum_{j=1}^N A_{ij}s_j,
\qquad
u \;\equiv\; \min_{1\le i\le N} S_i.
\label{eq:def_Si_u_appendix}
\end{equation}
For each fixed $i$, $S_i$ is a linear combination of independent Gaussians, hence Gaussian.
Moreover,
\begin{equation}
\mathbb E[S_i]=0,
\qquad
\mathrm{Var}(S_i)=\frac{1}{N}\sum_{j=1}^N \mathrm{Var}(A_{ij})=\delta_{ii},
\end{equation}
so that
\begin{equation}
S_i \sim \mathcal N(0,1).
\label{eq:Si_standard_normal}
\end{equation}

Since different rows of $\mathbf{A}$ are independent, $\{S_i\}_{i=1}^N$  are independent as well.
Therefore, for any stable fixed point $\vec{s}$, the random variable $u$ is the minimum of $N$
i.i.d. standard normal variables.

\subsubsection{Exact finite-\(N\) distribution of \(u\)}
Let $\Phi$ and $\varphi$ denote the CDF and PDF of the standard normal.
By independence, the probability of the minimum in a sample of $N$ 
i.i.d. random variables from a standard normal is
\begin{equation}
\mathbf P(u < x)
= 1-\big(1-\Phi(x)\big)^N.
\label{eq:survival_u_exact}
\end{equation}
Hence, the CDF and PDF of $u$ are
\begin{align}
F_N(u) &= \mathbf P(u\le x)= 1-\big(1-\Phi(u)\big)^N, \label{eq:cdf_u_exact}\\
\rho_N(u) &= F_N'(u)= N\,\varphi(x)\,\big(1-\Phi(x)\big)^{N-1}.
\label{eq:pdf_u_exact}
\end{align}
These are standard order-statistics identities.

%\subsection{Exponential growth of SFPs for symmetric coupling matrices}
%\label{app:exp_gro}
%In this appendix we want to provide a quantitative appreciation of the 
%basic difference
%between asymmetric and symmetric matrices just by reporting  
%an example of the exponential growth of the number of SFPs of model 
%(\ref{dyneq}) with  symmetric coupling matrices for quite a large value
%of the coupling, namely $g=5$ (see Fig. \ref{fig:sfp_vs_N}). 
%\begin{figure}[t]
%  \centering
%  \includegraphics[width=1\linewidth]%{Images_Final_Version/numero_punti_fissi_vs_N.png}
%  \caption{
%           Number of stable fixed points (SFP) as a function of the system size $N$,
%           shown on a linear--log scale. Blue markers denote the mean SFP averaged over
%           disorder realizations, with error bars. The red line is a least-squares fit that confirms an exponential growth
%           $\mathrm{SFP} \sim e^{0.159 N}$.}
%  \label{fig:sfp_vs_N}
%\end{figure}
%For asymmetric coupling matrices of the same size, for $g=5$ the measure of the
%basins of attraction of SFPs reduces so much as to make any numerical estimate %of
%their number practically unaccessible. 

\subsection{Symmetric Matrices}
\label{symmatr}
%\subsubsection{Setup and notation}
Here, we sketch the extension to the case of symmetric matrices of the perturbative criterion discussed in the manuscript for asymmetric ones.

Let $\mathbf{A}$ be an $N\times N$ symmetric Gaussian matrix with off-diagonal
entries $A_{ij}=A_{ji}\sim\mathcal{N}(0,1)$ ($i\neq j$) and diagonal variance
$d=\text{Var}(A_{ii})$. The two conventions of interest are $d=0\ (A_{ii}=0)$, and
$d=1\ (A_{ii}\sim\mathcal{N}(0,1))$. %, and $d=2\ \bigl(\mathbf{A}\to(\mathbf{A}+\mathbf{A}^{T})/\sqrt{2}\bigr)$.
%
%For a fixed sign pattern $\vec{s}\in\{\pm1\}^{N}$ define, as in the
%manuscript,
%\begin{equation}
%  S_i=\frac{1}{\sqrt{N}}\sum_{j=1}^{N}A_{ij}s_j,\qquad
%  u_i=s_iS_i,\qquad
%  u=\min_{1\le i\le N}u_i.
%  \label{eq:def}
%\end{equation}
%Throughout, $\varphi$ and $\Phi$ denote the standard normal PDF and CDF,
%and
%\begin{equation}
%  \rho_N(m)=N\,\varphi(m)\bigl(1-\Phi(m)\bigr)^{N-1}
%  \label{eq:rhoN}
%\end{equation}
%is the density of the minimum $m_N$ of $N$ i.i.d.\ standard normals.

% ==================================================================
%\subsubsection{Gauge invariance}
For a stable fixed point at $g=0$ we set $\vec{s}\in\{\pm1\}^N$.
Let $\mathbf{D}_s=\text{diag}(s_1,\dots,s_N)$ and $\tilde{\mathbf{A}}=\mathbf{D}_s\,\mathbf{A}\,\mathbf{D}_s$, i.e.\
$\tilde A_{ij}=s_iA_{ij}s_j$. The map $\mathbf{A}\mapsto\tilde{\mathbf{A}}$ leaves the
symmetric Gaussian ensemble invariant (gauge invariance), and we have
\begin{equation}
  u_i=s_iS_i=\frac{1}{\sqrt{N}}\sum_{j}\tilde A_{ij}.
  \label{eq:gauge}
\end{equation}
Hence, the joint law of $\{u_i\}$ does not depend on the pattern $\vec{s}$,
which also justifies averaging over all $2^{N}$ patterns.

% ==================================================================
%\subsubsection{Exact covariance}

In this case, the only source of correlation is the symmetry constraint
$\tilde A_{ik}=\tilde A_{ki}$. For $i\neq k$, the double sum
$\text{Cov}(u_i,u_k)=\frac{1}{N}\sum_{j,l}\mathbb{E}[\tilde A_{ij}\tilde A_{kl}]$
receives a single nonvanishing contribution ($j=k,\ l=i$), so that 
\begin{equation}
\begin{split}
  &\text{Cov}(u_i,u_k)=C_{ij}=\frac{1}{N}\quad(i\neq k),\qquad \\
  &\text{Var}(u_i)=C_{ii}=\frac{(N-1)\cdot 1+d}{N}=1+\frac{d-1}{N}.
\end{split}
  \label{eq:cov}
\end{equation}

% ==================================================================
%\subsubsection{Rank-one decomposition and law of the minimum}
In a more compact form we can write
%Eq.~\eqref{eq:cov} states that the covariance matrix is
\begin{equation}
  C_{ij}=\sigma_N^{2}\,\delta_{ij}+\frac{1}{N},\qquad
  \sigma_N^{2}=1+\frac{d-2}{N},
  \label{eq:C}
\end{equation}
which admits the exact representation
$u_i=\sigma_N v_i+z/\sqrt{N}$ with $v_1,\dots,v_N,z$ i.i.d.\ $\mathcal{N}(0,1)$.
Since the common shift $z/\sqrt{N}$ factors out of the minimum $m_N = \min_i v_i$, we have
\begin{equation}
  u\stackrel{d}{=}\sigma_N\,m_N+\frac{z}{\sqrt{N}},\qquad z\perp m_N.
  \label{eq:law}
\end{equation}
%where $m_N=\min{g_i}$.

% ==================================================================
%\subsubsection{Exact density}

By convolving Eq. \eqref{eq:law}, we obtain the probability density function
(PDF)
\begin{equation}
%\begin{split}
  \rho_N^{\mathrm{sym}}(u)=\int_{-\infty}^{+\infty}\!dz\,
  \sqrt{\frac{N}{2\pi}}\,e^{-Nz^{2}/2}\,
  \frac{1}{\sigma_N}\,\rho_N\!\left(\frac{u-z}{\sigma_N}\right).\\
%  & \sigma_N^{2}=1+\frac{d-2}{N}.
%\end{split}
  \label{eq:density}
\end{equation}
The corresponding cumulative density function (CDF) reads 
\begin{equation}
  P(u\le x)=1-\int_{-\infty}^{+\infty}\!dz\,
  \sqrt{\frac{N}{2\pi}}\,e^{-Nz^{2}/2}\,
  \left(1-\Phi\!\left(\frac{x-z}{\sigma_N}\right)\right)^{N}.
  \label{eq:cdf}
\end{equation}

%\paragraph{Zero-diagonal case ($d=0$).}
%Here $\sigma_N^{2}=1-2/N$, i.e.
%\begin{equation}
%  u=\sqrt{1-\frac{2}{N}}\;m_N+\frac{z}{\sqrt{N}}.
%  \label{eq:d0}
%\end{equation}
%At $N=20$: mean $=-1.7716$, standard deviation $=0.5460$ (Monte Carlo,
%$4\times10^{5}$ samples: $-1.7695$ and $0.5443$). Note that
%$\sigma_N^{2}+1/N=1-1/N<1$: the far left tail decays as
%$e^{-u^{2}/(2(1-1/N))}$, slightly below Eq.~\eqref{eq:rhoN} (about
%$-0.92$ nat at $u=-6$, $N=20$), while the right flank is strongly
%enhanced by the common mode ($+3.60$ nat at $u=0$). For $d=1$ the
%identity $\sigma_N^{2}+1/N=1$ makes the left tail coincide exactly with
%Eq.~\eqref{eq:rhoN}; for $d=2$ the tail lies above it.

% ==================================================================
%\subsubsection{Large-deviation rate}

\begin{figure}[t]
  \centering
  \includegraphics[width=1\linewidth]{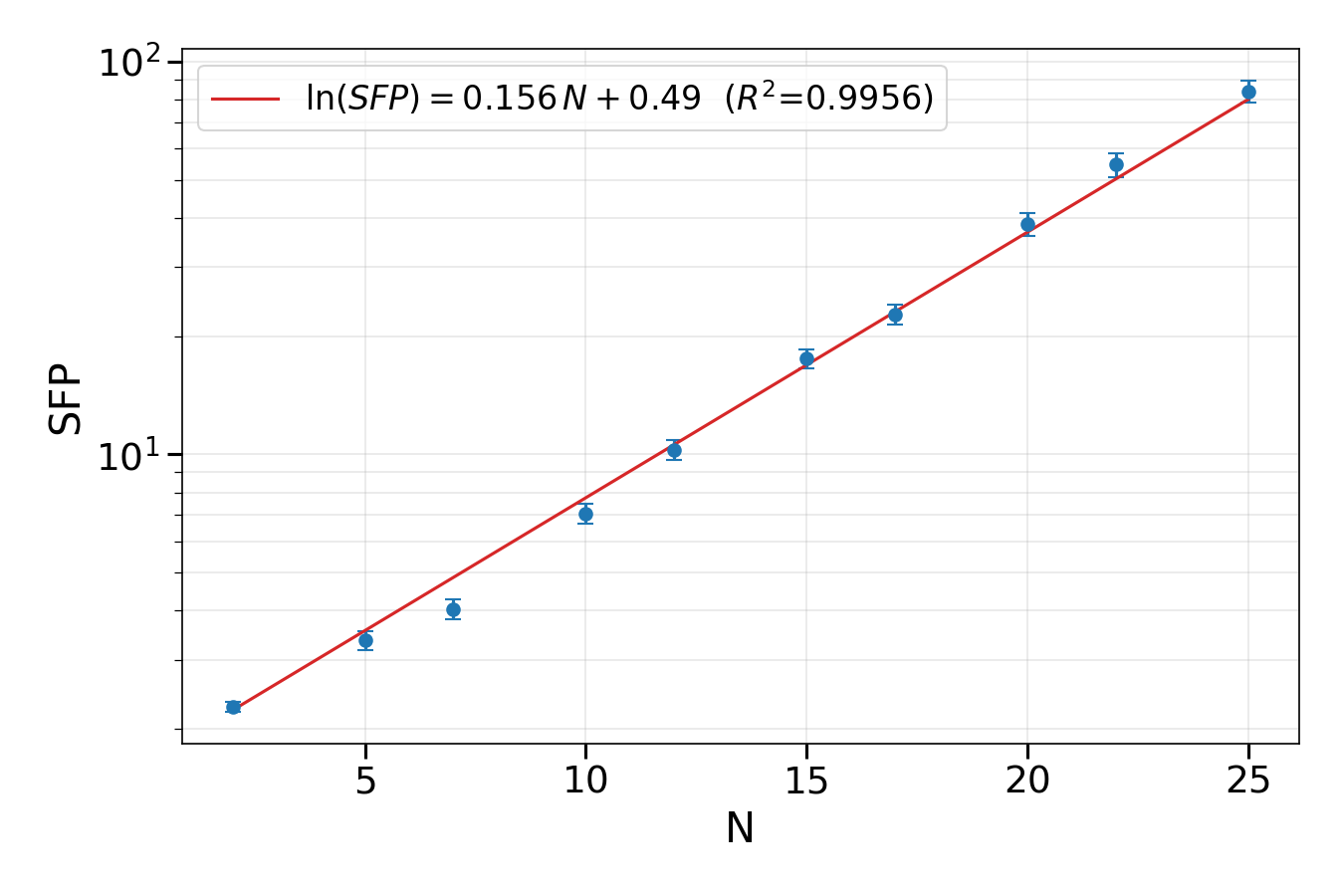}
  \caption{
           Number of stable fixed points (SFP) as a function of the system size $N$,
           shown on a linear--log scale. Blue markers denote the mean SFP averaged over
           disorder realizations, with error bars. The red line is a least-squares fit that confirms an exponential growth
           $\mathrm{SFP} \sim e^{0.159 N}$.}
  \label{fig:sfp_vs_N}
\end{figure}

Since the dominant common-mode fluctuation is $z=c\sqrt{N}$, with $c=O(1)$,
the diagonal convention drops out and the growth rate of the expected
number of surviving SFPs,
$h(u)=\lim_{N\to \infty}\frac{1}{N}\ln\!\bigl(2^{N}\rho_N(u)\bigr)$, becomes
\begin{equation}
  h_{\mathrm{sym}}(u)=\ln 2+\max_{c\ge 0}
  \left[\ln\bigl(1-\Phi(u-c)\bigr)-\frac{c^{2}}{2}\right],
  \label{eq:LD}
\end{equation}
which coincides in the limit $u\to-\infty$ with the result $\ln[2(1-\Phi(u))]$ 
obtained for asymmetric matrices. On the other hand, at variance with the 
asymmetric case, $h_{\mathrm{sym}}(u)$ does not vanish at $ u = 0$ (see the inset of Fig. \ref{fig:scaling} in the manuscript):
\begin{equation}
\begin{split}
  &h_{\mathrm{sym}}(0)=\ln 2+\ln\Phi(c^{*})-\frac{c^{*2}}{2}=0.1992\ldots,\\
  &\varphi(c^{*})=c^{*}\Phi(c^{*})\ \Rightarrow\ c^{*}\simeq0.5061.
\end{split}
  \label{eq:h0}
\end{equation}
Note that these analytic formulae hold in the limit $N\to\infty$. In order
to appreciate the reliability of the perturbative criterion,
%infuence of finite size effects 
in Fig.~\ref{fig:sfp_vs_N} we
report an example of the exponential growth of the number of SFPs of model 
(\ref{dyneq}) with  symmetric coupling matrices for 
%quite a large value
%of the coupling, namely 
$g=5$, i.e. a coupling value quite far from the perturbative regime.
Taking into account that this numerical estimate of the growth rate,
0.159, is unavoidably affected by hard-to-quantify finite size effects, it
is remarkable its closeness to the theoretical prediction 0.1992.
%thus confirming the
%effectiveness of the adopted perturbative criterion also for symmetric %coupling matrices.

\subsection{Large-$g$ limit}
\label{app:largeg}

In order to analyze dynamics (\ref{dyneq}) in the limit of large values of $g$ it is worth rescaling $x_i \to \sqrt{g} x_i$ and $t \to g t$, thus
obtaining
\begin{equation}
\label{dyneq_L_g}
\frac{d x_i }{dt} = - x_i(x_i^2 -\frac{1}{g}) + \frac{1}{\sqrt{N}} \, \sum_{j = 1}^{N} \, A_{ij} x_j, \, i =1,2, \cdots, N
\, 
\end{equation}
In the limit of $g \to +\infty$, while inducing also the asymptotic limit $t \to  +\infty$, this set of equations simplifies to
\begin{equation}
\label{dyneq_L_inf}
\frac{d x_i }{dt} = - x_i^3 + \frac{1}{\sqrt{N}} \, \sum_{j = 1}^{N} \, A_{ij} x_j, \, i =1,2, \cdots, N
\, 
\end{equation}
If the coupling matrix $A_{ij}$ would be diagonal with real eigenvalues
$\{ \lambda_i \}_{i=1}^N$ these equations represent a set of $N$ Duffing
oscillators, with stable fixed point in $0$ if $\lambda_i < 0$ or in
$\pm \sqrt{\lambda_i}$ if $\lambda_i > 0$. In the case of asymmetric
coupling matrices we have found numerically that dynamics (\ref{dyneq_L_inf})
yields an asymptotic evolution converging to a low-dimensional  chaotic attractor (data not shown).

\end{document}

%% file: preamble.tex
\usepackage{amsthm}
\usepackage{mathtools}
\usepackage{physics}
\usepackage{xcolor}
\usepackage{graphicx}
\usepackage[left=23mm,right=13mm,top=35mm,columnsep=15pt]{geometry} 
\usepackage{adjustbox}
\usepackage{placeins}
\usepackage[T1]{fontenc}
\usepackage{lipsum}
\usepackage{csquotes}